\documentclass[twocolumn,pra,aps, showpacs]{revtex4}

\usepackage{amsmath}
\usepackage{amssymb}
\usepackage[dvips]{graphicx}
\usepackage{psfrag}

\newcommand{\qed}{\hspace*{\fill}$\square$}

 \newcommand{\C}{\mathbf{C}}
 \newcommand{\N}{\mathbf{N}}
 \newcommand{\Z}{\mathbf{Z}}

 \newcommand{\funcion}[3]{#1:\,#2\longrightarrow #3}

 \newcommand{\vect}[1]{\boldsymbol{\mathrm{#1}}}
 
 \newcommand{\sset}[1]{ \{#1\} }
 \newcommand{\half}{\frac 1 2}

 \newcommand{\prima}{^\prime}
 \newcommand{\primas}{^{\prime\prime}}
 
 \newcommand{\nin}{ \not\in}

 \newcommand{\ket}[1]{|#1\rangle}
 \newcommand{\bra}[1]{\langle #1|}
 
 \newcommand{\ketbradif}[2]{\ket{#1}\bra{#2}}
 \newcommand{\ketbra}[1]{\ketbradif {#1}{#1}}

 \newcommand{\hilb} {H}
 \newcommand{\pauli} {\mathcal P}
 \newcommand{\stab} {\mathcal S}
 \newcommand{\gauge} {\mathcal G}
 \newcommand{\code} {C}
 \newcommand{\correction} {\mathcal R}
 \newcommand{\channel} {\mathcal E}
 
 \newcommand{\ban} {L}
 \newcommand{\suba} {A}
 \newcommand{\subb} {B}

 \newcommand{\rr}{\mathrm{r}}
 \newcommand{\rg}{\mathrm{g}}
 \newcommand{\rb}{\mathrm{b}}
 \newcommand{\rc}{\mathrm{c}}

\renewcommand{\b}[1]{\mathbf{#1}}
\newcommand{\id}{\b 1}

\def\be{\begin{equation}}
\def\ee{\end{equation}}

\begin{document}

\title{Topological Subsystem Codes}
\author{H. Bombin}
\affiliation{Department of Physics, Massachusetts Institute of Technology, Cambridge, MA, 02139 USA\\
Perimeter Institute for Theoretical Physics, 31 Caroline St. N., Waterloo, Ontario N2L 2Y5, Canada}

\begin{abstract} We introduce a family of 2D topological subsystem quantum error-correcting codes. The gauge group is generated by 2-local Pauli operators, so that 2-local measurements are enough to recover the error syndrome.  We study the computational power of code deformation in these codes, and show that boundaries cannot be introduced in the usual way. In addition, we give a general mapping connecting suitable classical statistical mechanical models to optimal error correction in subsystem stabilizer codes that suffer from depolarizing noise.
\end{abstract}


\maketitle

\section{Introduction}

Quantum error correction \cite{Shor_QEC, Steane_QEC, Knill_QEC, Bennett_QEC} and fault-tolerant quantum computation \cite{Shor_FTQC, Knill_FTQC, Aharonov_FTQC, Gottesman_FTQC, Preskill_FTQC} promise to allow almost perfect storage, transmission and manipulation of quantum information. Without them, quantum information processing would be doomed to failure due to the decoherence produced by interactions with the environment and the unavoidable inaccuracies of quantum operations.

The key concept in quantum error correction is the notion of quantum code. This is a subspace of a given quantum system where quantum information can be safely encoded, in the sense that the adverse effects of noise can be erased through an error correction procedure. In practice this procedure is also subject to errors and thus it should be as simple as possible to minimize them. Naturally, the meaning of `simple' will depend on particular implementations. A common situation is that interactions are restricted to quantum subsystems that are close to each other in space. In those cases, the locality of the operations involved in error correction becomes crucial.

The stabilizer formalism \cite{Gottesman_stabilizer, Calderbank_stabilizer} provides a unified framework for many quantum codes. In stabilizer codes the main step for error correction is the measurement of certain operators, which may be local or not. A class of codes where these measurements are intrinsically local is that of topological stabilizer codes \cite{Kitaev_nonAbelian, Dennis_TQM, Bombin_CC2d, Bombin_CC3d}. In a different direction, locality can also be enhanced by considering more generally stabilizer subsystem codes \cite{Bacon_3d, Poulin_subsystem}. The present work provides an example of a family of codes which can be labeled both as `topological' and `subsystem'. 

Topological codes where originally introduced with the goal of obtaining a self-protecting quantum computer \cite{Kitaev_nonAbelian}. This idea faces important difficulties in low dimensions, since thermal instabilities are known to occur \cite{Dennis_TQM, Alicki_stability2d, Alicki_stability4d}. On the other hand, topological codes are local in a natural way and have very interesting features in the context of active error correction. For example, they do not only allow to perform operations transversally \cite{Bombin_CC2d, Bombin_CC3d} but also through code deformations \cite{Dennis_TQM, Bombin_deformation}. Moreover, there exist a useful connection between error correction in topological codes and certain classical statistical models \cite{Dennis_TQM, Wang_topoStat, Katzgraber_3body}. 

Stabilizer subsystem codes are the result of applying the stabilizer formalism to operator quantum error correction \cite{Kribs_OQEC}. In subsystem codes part of the logical qubits that form the code subspace are no longer considered as such but, rather, as gauge qubits where no information is encoded. This not only allows the gauge qubits to absorb the effect of errors, but has interesting consequences for error correction. It may allow to break up each of the needed measurements in several ones that involve a smaller number of qubits \cite{Bacon_3d, Poulin_subsystem}. An example of this is offered by Bacon-Shor codes \cite{Bacon_3d}, in which the basic operators to be measured can have support on an arbitrarily large number of qubits, yet their eigenvalues can be recovered from 2-local measurements that do not damage encoded information. Moreover, the pairs of qubits to be measured together are always neighbors in a 2D lattice. Thus, subsystem codes can have very nice locality properties.

The 2D topological subsystem codes introduced here show all the characteristic properties of topological codes and at the same time take profit of the advantages of subsystem codes. Some of them are:
\begin{itemize}
\item The codes are local in a 2D setting, which can be flat.
\item The measurements needed for error correction only involve two neighboring qubits at a time, as in Bacon-Shor codes. This is an important advantage with respect to other topological codes, such as surface codes, that require measuring groups of at least 4 qubits. 
\item Most errors of length up to $cn$ are correctable, where $c$ is some constant and $n$ the number of physical qubits. This feature, common to topological codes, follows from the fact that logical operators are related to strings that wind nontrivially around the surface where the code is defined. 
\item Error correction must be done only `up to homology', an important simplification that allows the introduction of specific tools.
\item One can naturally perform certain logical gates through `deformations' of the code. This feature, however, is less powerful that in other topological codes because boundaries cannot be introduced in the usual way.
\end{itemize}

Since these codes are topological, it is natural to expect a connection between their correction procedures and suitable classical statistical models. However, the mapping between surface codes and random Ising models \cite{Dennis_TQM} makes stronge use of their CSS structure \cite{Calderbank_CSS, Steane_CSS}, and the same is true for the one between color codes and 3-body random Ising models \cite{Katzgraber_3body}. The CSS structure makes it possible to completely separate the correction of phase flip and bit flip errors, making the problem classical and enabling the connection. Indeed, there exist similar mappings from classical codes to statistical models \cite{Nishimori_statInfo}. Fortunately, as explained in section VI, the approach can be generalized even in the absence of this separation. Moreover, the subsystem structure is also compatible with the approach, so that it can be applied to the family of codes of interest.

The paper is organized as follows. Sections II and III go over several aspects of quantum error correction and topological codes, respectively, setting up a framework for the rest of the paper. Section IV introduces the family of topological subsystem codes and presents a thorough study of their properties. Section V offers the construction of a general mapping between error correction in subsystem codes and classical statistical models. Section VI is devoted to conclusions.

\section{Stabilizer quantum error-correcting codes}
\label{sec:subsystem}

This section summarizes the notions of quantum error correction that will be needed in the rest of the paper. It mainly reviews stabilizer codes, both in the subspace and the more general subsystem formulation. Ideal error correction procedures and their success probability are also considered.

\subsection{Quantum error correction}

Quantum error correction deals with the preservation of quantum information in the presence of noise. Both the noise $\channel$ and the error recovery $\correction$ are modeled as quantum operations or channels $\funcion { \channel, \correction }{\ban (\hilb)}{\ban (\hilb)}$, where $\ban(\hilb)$ is the space of linear operators on $\hilb$, the Hilbert space associated to the quantum system under consideration. Such maps can always be expressed in the operator-sum representation. For example, the noise is $\channel (\rho)=\sum_i E_i\rho E_i^\dagger$ for some $E_i\in\ban (\hilb)$, which will be denoted by $\channel=\sset{E_i}$.

In the original formulation of quantum error correction \cite{Shor_QEC, Steane_QEC, Knill_QEC, Bennett_QEC}, quantum information is encoded in a subspace of $\hilb$, the code subspace $\code\subset \hilb$. The system undergoes a noisy process $\channel$
and afterwards an error recovery operation $\correction$ is performed. Then, given a code $\code$, a noise source $\channel$ is said to be correctable if there exists a recovery operation $\correction$ such that $\correction\circ\channel (\rho)=\rho$ for any state $\rho \in \ban (\code)$.

More generally, in the operator quantum error  correction formalism \cite{Kribs_OQEC}, information is encoded in a subsystem $\suba$, with $\code =\suba\otimes\subb$. Whatever happens to subsystem $\subb$ is irrelevant. That is, error recovery is possible for a quantum channel $\channel$ if there exists a recovery operation $\correction$ such that for any $\rho^\suba\in \ban(\suba)$ and $\rho^\subb\in \ban(\subb)$ it gives $\correction\circ \channel (\rho^\suba\otimes\rho^\subb)=\rho^\suba\otimes\rho^{\prime\subb}$ for some arbitrary $\rho^{\prime\subb}$. 

The sufficient and necessary condition for the noise process $\channel =\sset{E_i}$ to be  correctable \cite{Knill_QEC, Bennett_QEC, Kribs_OQEC} is that $PE_i^\dagger E_j P=\id^\suba\otimes g_{ij}^B$ for every $i$ and $j$, with $P$ the projector onto the code subspace. When this condition holds, the set of errors $\sset{E_i}$ is said to be correctable. Since adding a linear combination of the $E_a$ to the set does not change correctability, it is natural to consider correctable sets of errors as linear subspaces and to choose the most convenient operator basis. Generally the quantum system is composed of $n$ qubits, $\hilb \simeq (\C^2)^{\otimes n}$, and error operators are chosen to be Pauli operators, elements of the Pauli group $\pauli_n :=\langle i\id, X_1,Z_1,\dots, X_n, Z_n \rangle$. Here $X_i$, $Z_i$ are as usual the Pauli operators on the $i$-th qubit, $X=\ket 0\bra 1+ \ket 1 \bra 0$, $Z=\ketbra 0 - \ketbra 1$ in the orthonormal basis $\sset{\ket 0, \ket 1}$.

Usually error models are such that errors which affect more qubits are less likely to happen. Then it makes sense to correct as many errors as possible among those that have support on (act nontrivially on) a smaller number of qubits. The weight $|E|$ of a Pauli operator $E\in \pauli_n$ is defined as the number of qubits that form its support. When a code can correct all Pauli errors $E$ with $|E|\leq r$ it is said to correct $r$ errors.

\subsection{Stabilizer subspace codes}

A formalism that has been particularly successful for the development of quantum codes is the stabilizer formalism \cite{Gottesman_stabilizer, Calderbank_stabilizer}, in which the code $\code$ is described in terms of an Abelian subgroup $\gauge=\langle G_j \rangle \subset \pauli_n$ such that $-1\nin \stab$. Take the generators $S_j$ to be independent and let $s$ be the rank of $\stab$. The $n$ qubit Hilbert space $H$ can be partitioned according to the eigenvalues of the $S_j$ into $2^s$ isomorphic orthogonal subspaces $H=\bigoplus_{\vect s} \code_{\vect s}$. Here $\vect s=(s_j)$ is the error syndrome, with $s_j=\pm 1$ the eigenvalue of $S_j$. By convention, the code subspace $\code$ is that with $s_j=1$ for all $j$. It has dimension $2^k$, with $k=n-s$ the number of encoded or logical qubits. The reason to call $\vect s$ the error syndrome is that in can be obtained by measuring the $S_j$ and then used to infer which errors have produced.

It is easy to introduce a Pauli group for the $k$ logical qubits. Let $N(\stab)$ be the normalizer of $\stab$ in $\pauli_n$. Its elements are the Pauli operators that map the subspaces $C_{\vect s}$ onto themselves, and the quotient group $N(\stab)/\stab$ is isomorphic to $\pauli_k$. The logical Pauli operators are then generated by $\hat X_1, \hat Z_1,\dots, \hat X_k,\hat Z_k \in N(\stab)$, some chosen representatives of the images of $X_1, Z_1, \dots,X_k,Z_k\in\pauli_k$ under a given isomorphism.

It is also possible to characterize a stabilizer code with a pair $(U,s)$, where $U$ is an automorphism of $\pauli_n$ and $s$ an integer, $0\leq s\leq n$. Let $\tilde X_i$, $\tilde Z_i$ denote the images of $X_i$, $Z_i$ trough $U$. Then the stabilizer is $\stab = \langle \tilde Z_1,\dots,\tilde Z_s \rangle$.  This approach directly provides a choice for encoded Pauli operators: $\hat X_1:= \tilde X_{s+1}, \hat Z_1:=\tilde Z_{s+1}, \dots, \hat X_k:=\tilde X_n, \hat Z_k:= \tilde Z_n$. 

Pauli errors $E$ have a specially simple effect on the encoded states, as they map the subspaces $C_{\vect s}$ one onto another. Set  $\vect s_E=\vect s$ when $ES_j E^\dagger= s_j S_j$. Then a Pauli error $E$ maps $\code$ onto $\code_{\vect s_E}$. Pauli operators are divided in three categories. The elements of $\pauli_n-N(\stab)$  map the code to other subspaces and are termed detectable errors, as their effect can be detected by measuring the operators $\sset {S_j}$. The elements of $N(\stab)-\stab\prima$, with $\stab':=\langle i\id \rangle \stab $, map in a nontrivial way the code to itself and are thus called undetectable errors. Finally, the elements of $\stab\prima$ have no effect on encoded states $\rho\in\ban (\code)$. The distance $d$ of the code is defined as the minimum weight among undetectable errors. It determines the number of corrected errors, which is $\lfloor (d-1)/2\rfloor$. A code of $n$ qubits that encodes $k$ qubits and has distance $d$ is denoted $[[n,k,d]]$.

\subsection{Stabilizer subsystem codes}

The stabilizer formalism can also be used in the context of operator quantum error correction \cite{Bacon_3d, Poulin_subsystem}. Instead of being characterized by a stabilizer group, subsystem stabilizer codes are almost determined by a subgroup $\gauge\subset \pauli_n$, called the gauge group, such that $i\id\in \gauge$. Almost because, in addition, a stabilizer group $\stab$ has to be chosen such that $\stab'$, as defined above, is the center of $\gauge$. There are different choices for $\stab$ because the sign of some of its generators can always be flipped. This amounts to different choices for $\code$ in the decomposition $H=\bigoplus_{\vect s} \code_{\vect s}$.

The idea after the introduction of the gauge group is that gauge operations should not affect the encoded information. This forces to identify states such as $\rho$ and $G_j\rho G_j^\dagger$ as equivalent, giving rise to a subsystem structure $C_{\vect s}=A_{\vect s} \otimes B_{\vect s}$. The decomposition is such that the gauge operators $G_j$ act trivially in the $A_{\vect s}$ subsystems and generate the full algebra of operators of the $B_{\vect s}$ subsystems. Set $C=A\otimes B$, with $A$ the logical subsystem where information is encoded and $B$ the gauge subsystem that absorbs the effect of gauge operations. Since $\gauge/\stab\simeq \pauli_r$ for some $r$, $B$ consists of $r$ qubits. Similarly, the Pauli operators for the $k$ logical qubits are recovered from the isomorphism $N(\gauge)/\stab\simeq \pauli_k$, and $k+r+s=n$.

It is also possible to characterize a stabilizer subsystem code as a triplet $(U,s,r)$, where $U$ is an automorphism of $\pauli_n$ and $s,r\geq 0$ are integers with $r+s \leq n$. Using the same notation as above, the stabilizer and gauge groups are $\stab = \langle \tilde Z_1,\dots,\tilde Z_s \rangle$ and $\gauge = \langle i\id, \tilde Z_1, \dots, \tilde Z_{s+r}, \tilde X_{s+1},\dots,\tilde X_{r}\rangle$. The chosen logical Pauli operators are $\hat X_1:= \tilde X_{s+r+1}, \hat Z_1:=\tilde Z_{s+r+1}, \dots, \hat X_k:=\tilde X_n, \hat Z_k:= \tilde Z_n$. 

In subsystem codes detectable Pauli errors are the elements of $\pauli_n-N(\stab)$ and undetectable ones are those in $N(\stab)-\gauge$. Undetectable errors are directly related to logical Pauli operators. Indeed, $N(\stab)/\gauge\simeq N(\gauge)/\stab\prima$, through the following correspondence. For any $E\in N(\stab)$ there exists a $G\in \gauge$ such that $EG\in N(\gauge)$, and if $G'\in \gauge$ is such that $EG\prima\in N(\gauge)$ then $GG\prima\in \stab\prima=\gauge\cap N(\gauge)$. The distance $d$ of the code is defined as for subspace codes and has the same implications regarding error correction. A subsystem code of $n$ qubits that encodes $k$ qubits and has $r$ gauge qubits and distance $d$ is denoted $[[n,k,r,d]]$.

\subsection{Syndrome measurement}

An interesting property of stabilizer subsystem codes is that they may allow an easier measurement of the stabilizer generators. This is so because it is possible to substitute the direct measurement of an stabilizer element $S$ by an indirect one, in which $t$ self-adjoint gauge operators $G_i$ such that $S=G_1\cdots G_t$ are measured. It may be the case that the $G_i$ have a smaller weight than $S$. For example, in the family of Bacon-Shor codes the gauge generators $G_i$ have always weight $|G_i|=2$ but the smallest stabilizer generators have an arbitrarily large weight. Such cases offer two important advantages. On the one hand, the smaller the weight of a Pauli operator, the simpler the operations needed to measure it. This is specially relevant in fault-tolerant quantum computing, where error correction is considered a faulty process in itself, because simpler operations imply less errors. Secondly, it may be possible to measure the $G_i$ in parallel, with the corresponding saving of time. This is again relevant for fault tolerance, where the ability to perform measurements faster entails less errors.

Since the $G_i$ need not commute the ordering of the measurements is relevant. In general, the ordered measurement of a collection of $t$ operators $E_1, \dots, E_t\in \pauli_n^\dagger$ yields the effective measurement of an abelian group of self-adjoint Pauli operators $\mathcal M\subset \mathcal Z$, with $\mathcal Z$ the center of $\langle-\id, E_1, \dots, E_t\rangle$. In particular, $\mathcal M=\mathcal N\cap\mathcal Z$ with $\mathcal N$ the abelian group of those self-adjoint Pauli operators with eigenvalues fixed by the sequence of measurements. $\mathcal N$ can be computed iteratively, since adding an additional measurement $E_{t+1}$ changes $\mathcal N$ to $\mathcal N\prima= \langle E_{t+1}\rangle \cdot \mathcal (\mathcal N\cap N(E_{t+1})) $

\subsection{Error correction}\label{sec:error_correction}

Even if a noisy channels $\channel$ is not correctable for a given subsystem code, there exists some probability of performing a successful error recovery. For example, if each of the physical qubits that compose a code undergoes a depolarizing channel $\sset{(1-3p)^\half\id, p^\half X, p^\half Y, p^\half Z}$, then the noise is certainly not correctable, but the success probability can still be close to one. This section quantifies this probability, which is of primary importance for topological codes \cite{Dennis_TQM}.

Some notation is needed here. Set as equivalent $E\sim E\prima$ those operators $E,E\prima\in\ban(\hilb)$ that have the same action up to gauge elements, $E= E\prima G$ for some $G\in \gauge$. The corresponding equivalence classes will be denoted $\bar E$. Let $\sset{D_i}_{i=1}^{4k}\subset N(\stab)$ be a particular set of representatives for $N(\stab)/\gauge$, taking in particular $D_1=\id$. The $D_i$ with $i> 1$ will represent the ways in which error correction can fail. For example, if there is one encoded qubit a choice is $\sset{\id, \hat X_1, \hat Z_1, \hat X_1\hat Z_1}$. To exted the equivalence of operators to channels, we choose the minimal equivalence relation such that $\sset{E_i}\sim\sset{E\prima_i}$ whenever $E_i\sim E_i\prima$ for all $i$.

Assume an error model in which Pauli errors $E\in \pauli_n$ occur with a given probability $p(E)$. That is, $\channel = \sset{p(E)^{\half} E}_{E\in\pauli_n}$. Errors with different phases will not be distinguished when discussing $p(E)$ because phases are irrelevant. Up to gauge operations the error channel is $\channel \sim \sset{p(\bar E)^{\half} E}_{\bar E\in\pauli_n/\gauge}$, where $E$ denotes any chosen element of $\bar E$ and $p(\bar E)=\sum_{G\in\gauge/\langle i\id \rangle} p(EG)$ is the probability for a given class of errors to happen. This makes already apparent that class probabilities $p(\bar E)$ are more important than individual error probabilities $p(E)$.

Error recovery starts with the measurement of the stabilizer generators $S_j$. This yields the error syndrome $\vect s$, which limits the errors $E$ that have possibly happened to those with $\vect s_E=\vect s$. These possible errors are arranged into different classes, which may be labeled by choosing any possible error $E$ and taking as representatives the elements $\sset{ED_i}$. Then the conditional probability for the class of errors $\bar E$ to have happened given the syndrome outcome $\vect s$ is
\begin{equation} \label{conditional_probability}
p(\bar E|\vect s) = \frac{p(\bar E)}{\sum_i p(\bar E\bar D_i)}.
\end{equation}
Suppose that these conditional probabilities can be computed, which may be potentially difficult due to the combinatorics. Then the class $\bar E=\bar E_{\vect s}$ that maximizes $p(\bar E|\vect s)$ is known and the optimal recovery operation is $\correction = \sset{E_{\vect s} P_{\vect s}}_{\vect s}$,  where $P_{\vect s}$ is the projector onto the subspace $C_{\vect s}$. The combined effect of errors and recovery is $\correction\circ \channel\sim \sset{p_i^\half D_i}_i$ for some probabilities $p_i$ that only depend on the error distribution $p(\bar E)$. This gives a success probability for the error recovery
\begin{equation} \label{success}
p_0=\sum_{\vect s} p(\bar E_{\vect s}).
\end{equation}

A bad feature of the expression \eqref{success} is that it depends on $E_{\vect s}$. Consider an alternative error correction procedure where the class of errors with the maximum probability $\bar E_{\vect s}$ is not always chosen. Instead, an operator from a class $\bar E$ is applied with probability $p(\bar E| \vect s)$, so that $\correction\prima = \sset{p(\bar E|\vect s)^\half  E P_{\vect s}}_{\vect s, \bar E}$. The success probability for this randomized correction procedure is
\begin{equation} \label{success_prime}
p\prima_0=\sum_{\bar E} p(\bar E) \,p(\bar E | \vect s_E)=\sum_E p(E) \,p(\bar E | \vect s_E).
\end{equation}
This procedure is at best as successful as the original one, giving the bound $p_0\prima\leq p_0$. Notice that $p_0=1$ if and only if $p_0\prima=1$. It follows that $p_0=1$ if and only if for any $D\in N(\stab)-\gauge$
\begin{equation} \label{success_condition}
\sum_{E} p(E) \,p(\bar E\bar D | \vect s_E)=0.
\end{equation}
This was the condition used in \cite{Dennis_TQM} to characterize successful recovery.

\section{Topological stabilizer codes}

This section gathers together several aspects of topological stabilizer codes to provide a reference for section \ref{sec:subsystem_topological_codes}. The goal is to put the subsystem codes introduced there in a broader context, making apparent the similarities and differences with previously known local and topological codes.

\subsection{Local codes}

In the context of fault-tolerant quantum computing it is advantageous to be able to perform the syndrome measurements in a simple way. It may be desirable, for example, that the number of qubits that form the support of the operators to be measured is small. Similarly, it may be convenient that the qubits that form the code only belong to the support of a small number of such operators. This ideas can be formalized to give rise to the notion of local families of codes \cite{Kitaev_nonAbelian}. In particular, a family of stabilizer subspace codes $\sset{\code_i}$ is local when i/ it contains codes of arbitrary distance and ii/ there exist two positive integers $\mu, \nu$ such that for each $\code_i$ there exists a family of generators of the stabilizer ${S_j}$ such that ii.a/ $|S_j|\leq \mu$ and ii.b/ the number of $S_j$-s with support on any given qubit is smaller than $\nu$. 

The drawback of such an abstract notion of locality is that it is not related in any way to a particular geometry. Many physical settings have qubits disposed in 1,2 or 3 spatial dimensions and only allow the direct interaction of nearby qubits. To reflect this fact, and without loss of generality regarding the nature of the lattices, the definition above may be modified as follows \cite{Bravyi_noGo}. First, the qubits of each code $\code_i$ are considered to be disposed in the vertices of some finite, periodic, cubic lattice of a given dimension $D$. Second, instead of ii/ there must exist a positive number $d$, independent of $\code_i$, such that the support of all the $S_j$ is contained in some cube containing $d^D$ vertices. A family of codes that is local in $D$ dimensions is always local in the previous sense.

A similar notion of local stabilizer subsystem codes may be defined by substituting the stabilizer with the gauge group. Notice that a family of subsystem codes might have local gauge generators but non-local stabilizer generators, as Bacon-Shor codes exemplify.

\subsection{Topological codes}\label{sec:topological_codes}

\begin{figure}
 \includegraphics[width=5cm]{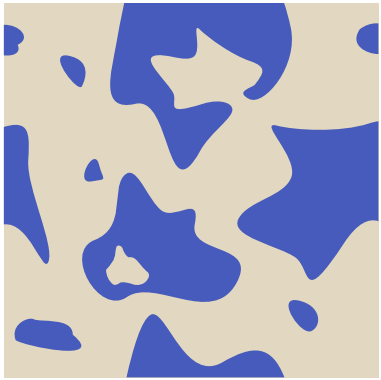}
 \caption{
In this figure the geometry is that of a torus, with opposite sides of the square identified. The dark region can be contained in a disconnected collection of simply connected regions, and thus cannot be the support of a non-detectable error in a topological code.}
 \label{fig:region}
\end{figure}

	Topological stabilizer codes are constructed from lattices on a given manifold, in such a way that the generators of the stabilizer are local with respect to the lattice \cite{Kitaev_nonAbelian, Bombin_homology, Bombin_CC2d, Bombin_CC3d}. Typically the set of qubits and the set of generators are directly related to sets of geometric objects such as the vertices, links or faces of the lattice. In order to distinguish truly topological codes from merely local ones, we propose the following criterium. \emph{In a topological stabilizer code, any operator $O\in N(\stab)$ with support in a subset of a region composed of disconnected pieces, each of them simply connected, has trivial action on logical qubits}. Stated this way, it is a rather vague criterium since no formal definition of region or connectedness is given. However, it will be enough for our purposes by adopting a reasonable interpretation when needed. Fig.~\ref{fig:region} shows an example of a region in a torus that cannot be the support of an undetectable error.

An enumeration of the common properties of known topological codes will be useful. First, the support of undetectable errors is topologically non-trivial in some well-defined way. Indeed, both in surface codes \cite{Kitaev_nonAbelian, Dennis_TQM, Bombin_homology} and color codes \cite{Bombin_CC2d, Bombin_CC3d} undetectable errors are related to homologically non-trivial cycles. Second, and closely related, the number of encoded qubits $k$ depends only upon the manifold in which the lattice is defined, and not the lattice itself. For example, for 2D surface and color codes the number of logical qubits is respectively $k=2-\chi$ and $k=4-2\chi$, with $\chi$ the Euler characteristic of the 2-manifold. Finally, an important property of topological codes is that their nature makes it possible to define them in many different lattices, typically rather arbitrary ones as long as some code dependent constraints are satisfied. In other words, topological codes display a huge flexibility. This is to be expected from constructions that only see the topology, not the geometry, of the manifolds to which they are related.

Notice that the locality of the stabilizer generators has been emphasized, with no mention to gauge generators. The reason is that, up to now, no genuinely subsystem topological codes have been known. This paper introduces a family of such codes. They have both local gauge generators and local stabilizer generators. Such locality properties should be expected from any topological subsystem code.

The family of Bacon-Shor codes \cite{Bacon_3d} provides an example of non-topological local gauge codes. These codes certainly do not satisfy the above criterium for any interpretation of connectedness that agrees with their 2D lattice geometry. Moreover, their geometry is completely rigid, in the sense that there is no clear way to generalize them to other lattices and manifolds.

It is interesting to observe that topological codes do not offer good $d/n$ ratios, which go to zero as larger codes are considered. For example, in two-dimensional surface or color codes $d=O(\sqrt n)$ holds (which is optimal among 2-dimensional local codes \cite{Bravyi_noGo}). But, as remarked in \cite{Kitaev_nonAbelian}, this is a misleading point because topological codes can correct most errors of order $O(n)$.

Finally, classical topological codes also exist \cite{Bombin_homology}. Unlike quantum ones, they can be obtained from mere graphs, that is, 1-dimensional objects.

\subsection{Topological quantum memories}\label{sec:topological_memory}

In \cite{Dennis_TQM} an interesting approach to the problem of indefinite preservation of quantum information was presented that makes use of topological codes. Since it will underly several discussions below, a brief summary is in order. The main idea is that information is encoded in a surface code and preserved by keeping track of errors. To this end, round after round of syndrome extractions must be performed. There are thus two sources of errors, since not only the code will suffer from storage errors but the stabilizer measurements themselves are also faulty. When the error rate is below certain threshold, the storage time can be made as long as desired by making the code larger, a feature that is only available in topological codes (for other codes one would have to use concatenation). Interestingly, this error threshold can be connected to a phase transition in a classical statistical model, a random 3D gauge model \cite{Dennis_TQM, Wang_topoStat}.

\subsection{Code deformation}

The flexibility of topological codes implies that they can be defined in many lattices. This feature makes natural the introduction of code deformations \cite{Dennis_TQM, Raussendorf_deformation, Bombin_deformation}, which are briefly described next. 

When two codes are very similar, in the sense that they differ only by a few qubits, it is possible to transform one onto another by manipulating these few qubits. This is specially natural for local codes that only differ locally. In particular, such local code deformations will not compromise the protection of the encoded information. These ideas where first explored in \cite{Dennis_TQM}, where the geometry of a surface code is transformed step by step to compensate a change in the lattice geometry provoked by a transversal gate. In \cite{Dennis_TQM} code deformations are also used to initialize the code with an arbitrary state. This is done by `growing' the code from a single qubit encoding the desired state. Notice that in this case encoded information is not protected on the early stages of the code deformation, when the code is still small.

In general \cite{Bombin_deformation}, two main kinds of deformations may be distinguished: those which change the number of encoded qubits and those which do not. The formers can be used to initialize and measure encoded qubits, and the latter to perform operations on encoded qubits. In the case of topological codes, code deformations amount to changes in the geometry of a lattice, which may ultimately be understood as changes in the geometry of a manifold. When the topology of the manifold changes, initialization or measurement of encoded qubits will happen in a well-defined way \cite{Bombin_deformation}. When the manifold undergoes a continuous transformation that maps it to itself, a unitary operation is performed on the encoded qubits \cite{Bombin_deformation}. This unitary operation only depends on the isotopy class of the mapping.

Code deformation can be naturally integrated with the successive rounds of stabilizer measurements mentioned in the previous section. In particular, as long as the deformations are local, one can perform them simply by changing the stabilizers to be measured at each stage of error detection \cite{Bombin_deformation}.

\subsection{String operators} \label{sec:string_operators}

In this and subsequent sections we only consider 2-dimensional topological codes with qubits as their basic elements, because the topological subsystem codes introduced in section IV fall into this category. For the same reason, subsystem code language, as opposed to subspace, will be used.

\begin{figure}
\psfrag{a}{$a$}
\psfrag{b}{$b$}
\psfrag{c}{$c$}
\psfrag{d}{$d$}
\psfrag{e}{$e$}
\psfrag{f}{$f$}
 \includegraphics[width=5cm]{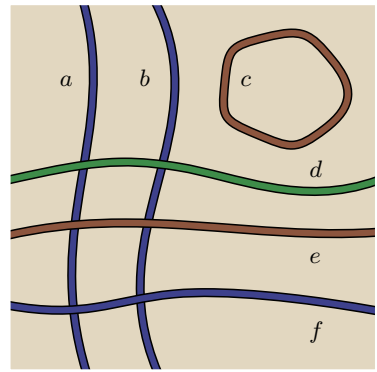}
 \caption{
In this figure the geometry is that of a torus, with opposite sides of the square identified. The colored curves represent the support of string operators, with color standing for strings labels. The strings $a$ and $b$ enclose a region and thus are homologically equivalent, producing equivalent operators. The string $c$ encloses a region and thus is homologically trivial, producing a stabilizer element. The strings $d$, $e$ and $f$ are homologically equivalent but have different labels, producing inequivalent operators.}
 \label{fig:strings}
\end{figure}

In known 2D topological codes the logical Pauli operators $\hat X_1, \dots, \hat Z_k\in \N(\gauge)$ can be chosen to be string operators \cite{Kitaev_nonAbelian, Bombin_CC2d}. These are operators $O_s$ with support along a set of qubits $s$ that resembles a closed string. There are several types of strings, labeled as $\sset{l_i}$. Two strings $s,s\prima$ of the same type that enclose a given region, like $a$, $b$ in Fig.~\ref{fig:strings}, give equivalent operators $O_sO_s\prima\in \stab\prima$. In other words, only the homology of the strings is relevant. In particular, boundary strings, those that enclose a region like $c$ in Fig.~\ref{fig:strings}, produce operators in $\stab\prima$. Moreover, $\mathcal S\prima$ is generated by boundary strings of some minimal regions. When two strings $s,s\prima$ cross once, like $a$ and $d$ in Fig.~\ref{fig:strings}, $O_s$ and $O_{s\prima}$ commute or not depending only on the labels of the strings. Finally, two strings $s$, $s\prima$ with a common homology class can be combined in a single string $s\primas$ of a suitable type, in the sense that $O_sO_{s\prima}O_{s\primas}\in \stab\prima$. For example, $d$ and $e$ in Fig.~\ref{fig:strings} can be combined in a string $f$ of a suitable type.

All this properties can be captured in a group $L\prima \simeq \pauli_t$, for some $t$ that depens on the code. For example, $t=1$ for surface codes and $t=2$ for color codes. The group $L:=L\prima/\langle i\id \rangle$ has as its elements the string types $\sset {l_i}$ and the product corresponds to string combination. The comutation rules are recovered from $L\prima$: crossing strings commute or not depending on whether their labels commute in $L\prima$. There are $2t$ labels $\bar l_1, \dots, \bar l_{t}$ that generate $L$. In a given manifold $2-\chi$ nontrivial cycles that generate the homology group can be chosen. Each of them gives rise to $2t$ string operators with labels $l_i$, and the total $2t(2-\chi)$ string operators generate $N(\gauge)/\stab\prima$, so that the number of encoded qubits is $k= t(2-\chi)$.

Instead of strings, in general one can consider string-nets, where the strings meet at branching points \cite{Bombin_CC2d}. The allowed branchings are those in which the product of all the labels involved is trivial. String-nets do not play a significant role in closed manifolds, but can be essential when the manifold has boundaries \cite{Bombin_CC2d}.

Let us check the criterium for topological codes of section \ref{sec:topological_codes} using the string operator structure. To this end a notion of connectedness is needed, but this can be obtained from the local generators of $\mathcal S\prima$: two regions or sets of qubits are disconnected from each other if no local generator has support on both of them at the same time. Let $Q_O$ denote the support of an operator $O$. Then if $O\in N(\stab)$ and $Q_O=Q_1\sqcup Q_2$ with $Q_1$ and $Q_2$ disconnected it follows that $O=O_1O_2$ and $Q_1=Q_{O_1}$, $Q_2=Q_{O_2}$ for some $O_1,O_2\in N(\stab)$. As for simply connectedness, it is easier to introduce a wider notion of `trivial' region. A set of qubits $Q$ forms a trivial region when there exist string operators $O_i$ that generate $N(\gauge)/\stab\prima$ and such that $Q_{O_i}\cap Q=\emptyset$. If $O\in N(\stab)$ is such that $Q_O$ is a trivial region then $[O,O_i]=0$ for the corresponding string generators $O_i$ and thus $O\in\gauge$. The criterium is satisfied within this language: if $O\in N(\stab)$ and $Q_O=Q_1\sqcup \dots\sqcup Q_t$ with the $Q_i$ pairwise disconnected and each of the $Q_i$ a trivial region, then $O\in\gauge$.

\subsection{Anyons}

Topological codes can be described in terms of string operators because they describe ground states of topologically ordered quantum models, that is, systems with emergent abelian anyons \cite{Kitaev_nonAbelian}. Anyons are localized quasiparticles with unusual statistics. String operator represent quasiparticle processes, and their commutation rules are directly related to the topological interactions of the anyons. Moreover, when an open-ended string operator is applied to the ground state, a pair of anyons is created on the ends of the string. The labels of the created anyons are those of the string, so that string labels are also quasiparticle labels. 

From the perspective of the code quasiparticles correspond to error syndromes, signaling a chain of errors along the string \cite{Dennis_TQM}. Thus, keeping track of errors in a topological code, recall section \ref{sec:topological_memory}, amounts to keep track of the wordlines of these quasiparticles. Error correction will success if the wordlines are correctly guessed up to homology \cite{Dennis_TQM}.

\subsection{Boundaries}\label{sec:boundaries}

From a practical perspective, codes that are local in closed 2-manifold like a torus are not very convenient. Instead, one would prefer to have planar codes. Thus, a way to create nontrivial topologies in the plane is needed, and this is exactly what is gained by introducing boundaries.

In a given code, different types of boundaries are possible. To start with, one can always consider random, structureless boundaries. The introduction of such boundaries will typically produce many local encoded qubits along the boundary. But these qubits are unprotected, and thus essentially useless.

\begin{figure}
\psfrag{a}{$a$}
\psfrag{b}{$b$}
\psfrag{c}{$c$}
\psfrag{d}{$d$}
\psfrag{e}{$e$}
\psfrag{f}{$f$}
 \includegraphics[width=7cm]{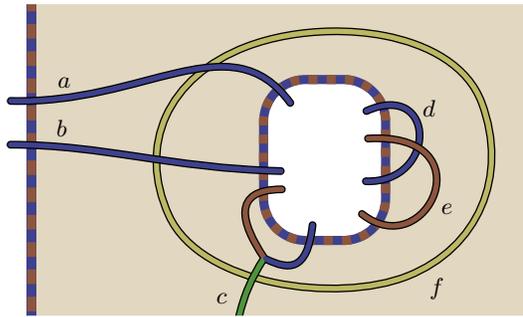}
 \caption{
This figure illustrates boundaries on 2D topological codes, which are displayed as dashed thick lines. The strings $a$ and $b$, together with the boundaries, enclose a region and thus produce equivalent operators. The string $c$ can end in the boundary because it can be decomposed in two strings that can end on it. The strings $d$ and $e$ enclose regions and thus produce stabilizer operators. The string $f$ produces an stabilizer element or an undetectable error, depending on wether its label is allowed in the boundary.}
 \label{fig:boundary}
\end{figure}

More interestingly, boundaries with well-defined properties and non-local encoded qubits are also possible \cite{Dennis_TQM, Bombin_CC2d}. The defining property of such boundaries is that strings $s$ with labels from certain subset $M\subset L$ are allowed to end in them, see Fig.~\ref{fig:boundary}, in the sense that $O_s$ belongs to the normalizer $N(\stab)$. In other words, the introduction of the boundary changes the notion of closed string, by allowing on the boundary loose ends of strings of suitable types. The notion of boundary string also changes. Two strings $s,s\prima$ of the same type that, together with boundaries in which they can end, form the the boundary of a given region, as $a$ and $b$ in Fig.~\ref{fig:boundary}, produce equivalent operators so that $O_s O_s\prima\in\stab\prima$. 

Notice that $M$ should be a subgroup of $L$, because if strings with labels $l,l\prima\in M$ can end in the boundary then so can strings with label $ll\prima$ by splitting before reaching the bondary, as string $c$ in Fig.  \ref{fig:boundary}. Also, any two labels $l,l\prima\in M$ must commute in $L\prima$. In other case, the stabilizer would contain anticommuting elements, which is not possible. This is illustrated by the strings $d$ and $e$ of Fig.~\ref{fig:boundary}, which must produce commuting operators. Finally, $L\prima$ should be maximal in the sense that for any $l\nin M$ there exist some $l\prima\in M$ such that $l$ and $l\prima$ anticommute in $L\prima$. In other case, according to the rules stated above, an $l$-string $s$ that surrounds a $M$-hole, like $f$ in Fig.  \ref{fig:boundary}, produces an operator $O_s$ that has to belong to $N(\stab)-\stab\prima$ because it is not a boundary but for which there is no other string $s\prima$ such that $\sset{O_s,O_s\prima}=0$. This is a contradiction.  It is in fact possible to relax this last maximality condition, but at the cost of getting a boundary between two topological codes, rather than a boundary between a code and the `vacuum'.

Remarkably, boundaries in topological codes are directly related to anyon condensation in the corresponding topologically ordered models \cite{Bombin_condensate}. It will become apparent in section \ref{sec:subsystem_boundaries} that this has important consequences, because only bosons can condense and this forbids certain types of boundaries.

\section{A family of topological subsystem codes}\label{sec:subsystem_topological_codes}

The subsystem codes introduced in this section have their origin in a spin-1/2 quantum model that shows topological order \cite{Bombin_2body}. The Hamiltonian of the model is a sum of 2-local Pauli terms that here will become the gauge generators. The Pauli symmetries of this Hamiltonian where already described in \cite{Bombin_2body}, and thus to some extent the codes where already implicitly considered in that work. Here we explicitly work out all the details from the code perspective. In addition, diverse aspects that are important in practice are explored, such as the possibility of introducing boundaries and the computational power of code deformations.

\subsection{Lattice and gauge group}\label{sec:subsystem_lattice}

\begin{figure}
\psfrag{(a)}{(a)}
\psfrag{(b)}{(b)}
 \includegraphics[width=8cm]{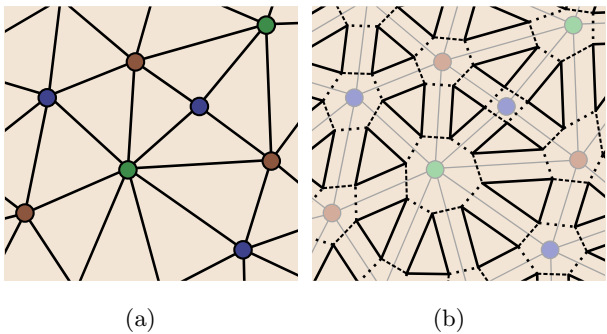}
 \caption{
(a) Part of a 2D lattice $\Lambda$ with triangular faces and 3-colorable vertices. (b) The lattice $\bar \Lambda$ derived from $\Lambda$. It is obtained by separating the triangles of $\Lambda$ and adding one face per edge and vertex of $\Lambda$. Its edges are classified in three types. Solid edges are $Z$-edges, dashed edges are $Y$-edges and dotted edges are $X$-edges. There is one qubit per vertex and the generators of the gauge group are related to edges. They are 2-local operators of the form $XX$, $YY$ or $ZZ$, depending on the edge type.}
 \label{fig:lattice}
\end{figure}

The family of codes $\code_\Lambda$ of interest is parametrized by tripartite triangulations $\Lambda$ of closed 2-manifolds, not necessarily orientable. That is, each code $\code_\Lambda$ is obtained from a 2-dimensional lattice $\Lambda$ such that (i) all faces $f\in F$ are triangular and (ii) the set of vertices $V$ can be separated in three disjoint sets, in such a way that no edge $e\in E$ connects two vertices from the same set. Fig.~\ref{fig:lattice}(a) shows an example. Alternatively, $\Lambda$ is the dual of a 2-colex \cite{Bombin_CCto}. Following the notation used in previous works, the three sets of vertices are colored as red, green and blue. The faces of $\Lambda$ will be simply called triangles. 

The first step in the construction of $\code_\Lambda$ is to derive a new lattice $\bar\Lambda$ from $\Lambda$, as exemplified in Fig.~\ref{fig:lattice}(b). In going from $\Lambda$ to $\bar\Lambda$, the triangles of $\Lambda$ separate from each other giving rise to new faces. In particular, each of the edges and vertices of $\Lambda$ contributes a face to $\bar\Lambda$. The edges of $\Lambda$ are divided into three subsets, $\bar E= \bar E_X\sqcup\bar E_Y\sqcup\bar E_Z$. In Fig.~\ref{fig:lattice}(b), $X$-edges are dotted, $Y$-edges are dashed and $Z$-edges are solid. The $Z$-edges form the triangles of $\bar\Lambda$. Each edge in $\Lambda$ contributes a $X$-edge and a $Y$-edge, in such a way that no two $X$-edges or two $Y$-edges meet. There are thus two ways to choose the sets of $X$ and $Y$ edges.

The definition of $\code_\Lambda$ is now at hand. First, physical qubits  correspond to the vertices of $\bar\Lambda$. Second, the gauge group is $\gauge_\Lambda:=\langle i \id\rangle\cdot \langle G_e\rangle_{e\in E}$, with generators $G_e$ related to the edges $e$ of $\bar\Lambda$. These take the form $G_e := \sigma_v\sigma_{v\prima}$ for $e\in \bar E_\sigma$, $\sigma = X,Y,Z$, where $v, v\prima$ are the vertices connected by $e$. Thus, the generators are 2-local.  This is an improvement with respect to previously known topological stabilizer codes, which have generators of weight at least 4.

\subsection{String operators}\label{sec:subsystem_string_operators}

This section describes $N(\gauge_\Lambda)$ and its center $\stab\prima_\Lambda= \gauge_\Lambda\cap N(\gauge_\Lambda)$ in terms of the string operator framework of section \ref{sec:string_operators}, which is valid for these codes. It turns out in particular that $L\prima\simeq\pauli_1$, as in surface codes. In section \ref{sec:subsystem_boundaries} it will be apparent, however, that there exist differences between the nature of the strings of $\code_\Lambda$ codes and those in surface codes. These are not captured by $L\prima$, which indeed does not contain all the information about the corresponding topological order. The details of the statements made in this section can be found in appendix  \ref{app:gauge}.

\begin{figure}
\psfrag{(a)}{(a)}
\psfrag{(b)}{(b)}
\psfrag{1}{$\id$}
\psfrag{X}{$X$}
\psfrag{Y}{$Y$}
\psfrag{Z}{$Z$}
 \includegraphics[width=8cm]{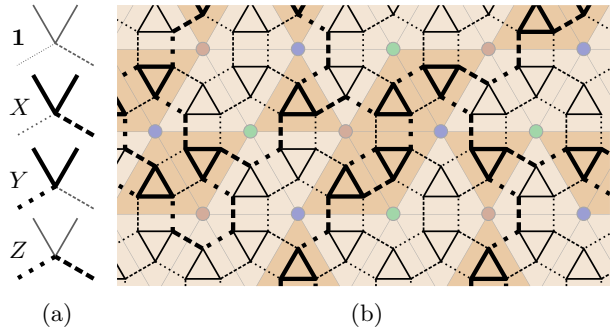}
 \caption{
(a) The four possible configurations at a given vertex for allowed subgraphs of $\bar\Lambda$. A different Pauli operator corresponds to each of them. (b) A subgraph $\gamma$ (thick lines) of a lattice $\bar\Lambda$ (thin lines), obtained from a regular triangular lattice $\Lambda$ (lightest lines).}
 \label{fig:normalizer}
\end{figure}

We first seek a graphical representation of $N(\gauge_\Lambda)$. Take any subgraph $\gamma$ of the graph of $\bar\Lambda$, such as the one in Fig.~\ref{fig:normalizer}(b), that has at each of its vertices one of the configurations of Fig.~\ref{fig:normalizer}(a). This graph $\gamma$ produces a Pauli operator $O_\gamma=\bigotimes_v \sigma_v$ with $\sigma_v=\id, X,Y,Z$ according to the correspondence of Fig.~\ref{fig:normalizer}(a). Observe that such operators $O_\gamma$ belong to $N(\gauge_\Lambda)$. Up to a phase, the correspondence between the elements of $N(\gauge_\Lambda)$ and graphs is one to one.

\begin{figure}
 \includegraphics[width=8cm]{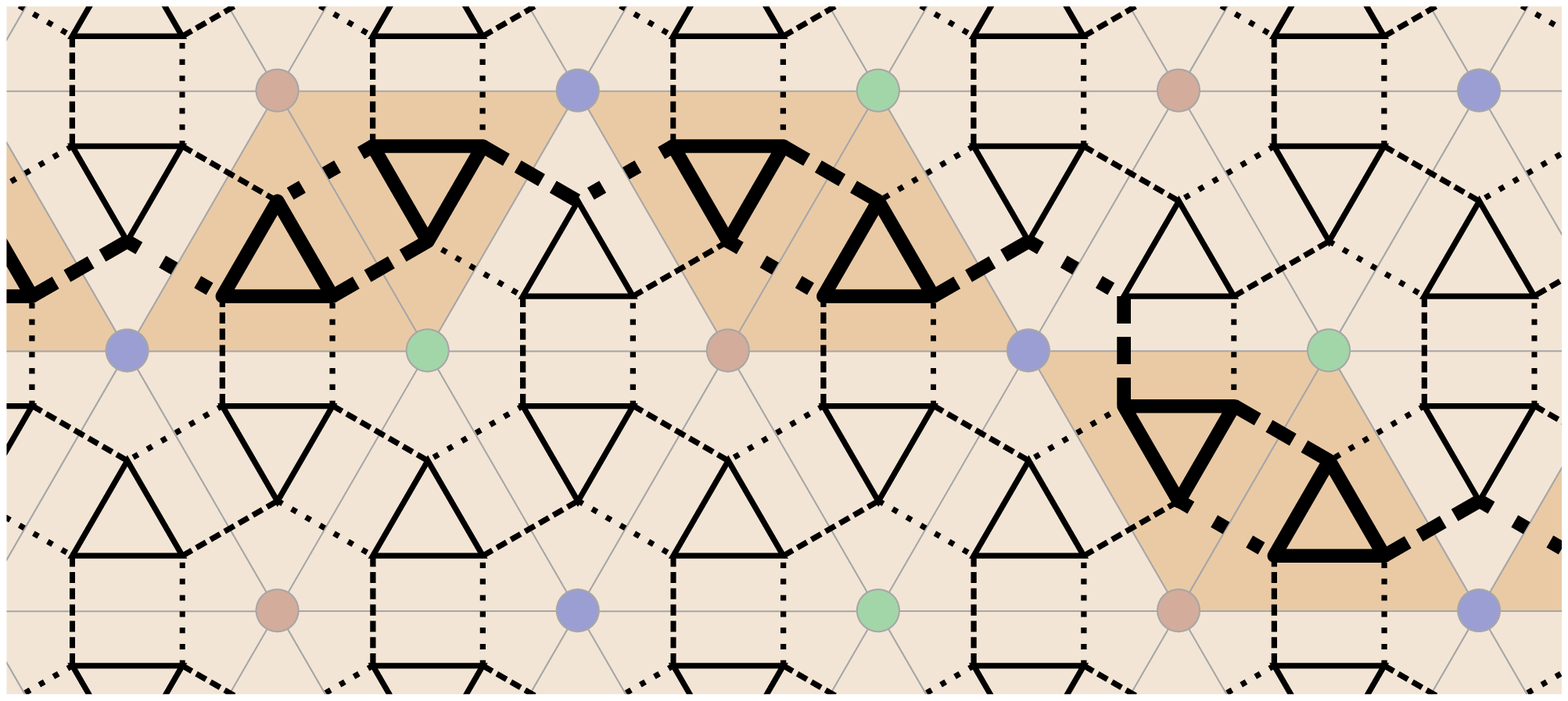}
 \caption{
A string operator as a subgraph $\gamma$ of $\bar\Lambda$, displayed in thick lines. Its triangles come in pairs, each of them connecting vertices of the same color in $\Lambda$.}
 \label{fig:string_ops}
\end{figure}

\begin{figure}
\psfrag{a}{a}
\psfrag{b}{b}
 \includegraphics[width=8cm]{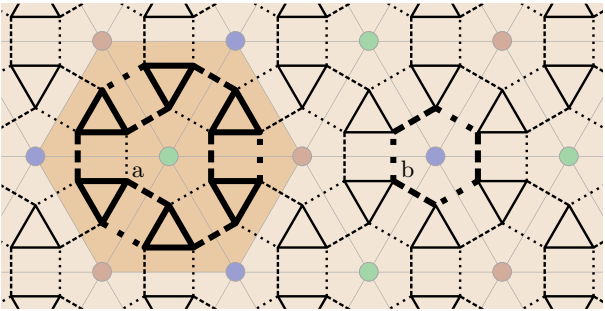}
 \caption{
Two examples of string operators $S_v^c$ related to vertices $v$ in $\Lambda$. The string $b$ shares color with the vertex that encloses, whereas for $a$ the two colors differ producing a more involved operator.}
 \label{fig:plaquette_ops}
\end{figure}

These graphs either contain all the edges of a triangle or none of them. Thus, each graph $\gamma$ determines a subset of triangles $T_\gamma$ of the original lattice $\Lambda$. In Figs.~\ref{fig:normalizer}-\ref{fig:plaquette_ops} this subset appears shaded. Notice that the number of triangles of $T_\gamma$ meeting at each vertex is even. In fact, any subset of triangles that meets this property can be realized as $T_\gamma$ for some $\gamma$.

String operators are obtained from string-like graphs such as the one in Fig.~\ref{fig:string_ops}. Notice in the figure how triangles can be paired in a specific way. These pairs of triangles connect always vertices of the same color from the original lattice $\Lambda$. These allows to classify strings accordingly with labels $\rr$, $\rg$, $\rb$. It is a simple exercise to check that crossing strings operators commute if they have the same color and anticommute in other case, in accordance with $L\prima$. String-nets can be formed by introducing branching points where three strings of different color meet.

The  group $\stab\prima_\Lambda$ is generated by small string operators related to vertices $v$ of the original lattice $\Lambda$. In particular, let us set $S_v^c=O_\gamma$, $c=\rr, \rg, \rb$,  with $\gamma$ the $c$-colored string  going around $v$, as shown in Fig.~\ref{fig:plaquette_ops}.  Then $\stab'_\Lambda=\langle i\id \rangle\cdot\langle S_v^c \rangle_{v,c}$. These generators are only subject to the relations 
\begin{equation}\label{constraints_stabilizer}
\prod_c S_v^c \propto \id,\qquad \prod_v S_v^c \propto \id, 
\end{equation}
where the first product runs over the three colors and the second over the vertices of $\Lambda$. As a consequence, the rank of $\stab_\Lambda$ is $s=2|V|-2$. Since the number of encoded qubits is $k=2-\chi$, it follows that the number of gauge qubits is $r=n-k-s= |3F|-2|V|+\chi=2|F|-\chi$, showing that gauge qubits see the global structure of the manifold.

What can string operators say about the code distance $d$? Given an operator $O_\gamma\in N(\gauge)$, consider the subset $E\prima$ of the edges of $\gamma$ with elements all the $X$- and $Y$-edges of $\gamma$ and one of the three $Z$-edges that correspond to each triangle in $T_\gamma$. Then $G=\prod_{e \in L} G_e\in\gauge_\Lambda$ and it is easy to check that $|O_\gamma G|=|T_\gamma|$. Therefore $d\leq d_T$, with $d_T$ the minimal length, in terms of the number of triangles, among nontrivial closed strings. A lower bound for $d$ is given in the next section.

\subsection{Homology of errors}

This section offers a homological description of error correction for $\code_\Lambda$. The main idea is that the error syndrome can be identified with the boundary of errors, considered as paths on the surface. Then error correction succeeds if this paths can be guessed up to homology. It is worth noting that the notation and results in this section will not be used again.

To fix notation, we recall first some basic notions. Let $\Delta$ denote the additive group of $\Z_2$ 1-chains in $\Lambda$. Its elements are sets of edges $\delta\subset E$ and addition is given by $\delta+\delta\prima= (\delta\cup \delta\prima) - (\delta\cap \delta\prima)$. The boundary $\partial \delta$ of $\delta\in \Delta$ is the set of vertices in which an odd number of edges from $\delta$ meet. The elements $\delta\in \Delta$ with $\partial \delta = 0$ are called cycles and form a subgroup $Z\subset \Delta$. Boundaries form a subgroup $B\subset Z$, generated by elements of the form $\delta=\sset {e_1,e_2,e_3}$ with $e_i$ the three edges of a given triangle. The first $\Z_2$ homology group of $\Lambda$ is $H_1:= Z/B\simeq \Z_2^h$ with $h=2-\chi$ the number of independent nontrivial cycles of the closed surface formed by $\Lambda$. Two chains $\delta,\delta\prima \in \Delta$ are said to be equivalent up to homology, $\delta\sim \delta\prima$, if $\delta+\delta\prima\in B$.

Consider a morphism $\funcion {f_\rr} {\pauli_n}{\Delta}$ defined by  $f_\rr (i)= \emptyset$ and the following action on single qubit operators $X_{\bar v}$, $Y_{\bar v}$, where $\bar v\in \bar V$. $X_{\bar v}$ anticommutes exactly with two operators of the form $S_v^\rr$, $v\in V$. The corresponding two vertices are connected by an edge $e\in E$, and $f_\rr (X_{\bar v}) = \sset e$. $f_\rr (Y_{\bar v})$ is defined analogously. It is easy to check that $f_\rr[\gauge_\Lambda] = B$ and that for any $O\in \pauli_n$ the set $\partial f_\rr(O)$ contains those vertices $v\in V$ such that $\sset{O,S_v^\rr}=0$. Moreover, if $\gamma$ is a string then $f_\rr(O_\gamma)\in Z$ and if $\gamma$ is red $f_\rr(O_\gamma)\in B$. Indeed, if $\sset{\gamma_i}$ is the set of red strings, then $f_\rr$ gives an isomorphism $N(\stab_\Lambda)/(\gauge_\Lambda\cdot\langle O_{\gamma_i}\rangle_i) \simeq H_1$.

Consider in addition an analogous morphism $f_\rb$ with blue color playing the same role as red in $f_\rr$. Then for any $O\in \pauli_n$ we have $O\in \gauge_\Lambda$ if and only if $f_c(O)\in B$ for $c=\rr, \rb$. Similarly, $O\in N(\stab_\Lambda)$ if and only if $f_c(O)\in Z$ for $c=\rr, \rb$. This shows that error correction will succeed as long as errors can be guessed up to homology. In detail, suppose that the code suffers a Pauli error $O$. The error syndrome can be expressed in terms of the two sets $\partial f_\rr(O), \partial f_\rb(O)\subset V$. Suppose that an attempt is made to correct the errors by applying some $O\prima \in \pauli_n$ such that $\partial f_c(O)=\partial f_c(O\prima)$ for $c=\rr, \rb$. Then error correction succeeds if and only if $O\prima O\in \gauge$, that is, if and only if $f_c(O)\sim f_c(O\prima)$ for $c=\rr, \rb$. 

Although error correction can be expressed in this homological terms, this is really not the most natural thing to do, basically because it involves an arbitrary choice of two of the three available colors. In this regard, notice that not any set of edges $\delta\in \Delta$ can be obtained from an operator $O\in \pauli_n$ as $\delta = f_\rr(O)$, and that the cardinalities of $f_\rr(O)$ and $f_\rb(O)$ by no means are enough to compute $|O|$. This makes unfeasible a direct translation of the ideas used in \cite{Dennis_TQM} for error correction in surface codes.

In order to give a lower bound for the distance $d$ of the code, the definition of the mappings $f_c$, $c=\rr,\rb$ must be modified. We set $f_c\prima(\bigotimes_{\bar v} \sigma_{\bar v}):=\sum_{\bar v} f_c\prima(\sigma_{\bar v})$, where $\sigma_{\bar v} =\id_{\bar v}, X_{\bar v},Y_{\bar v}, Z_{\bar v}$, and we fix $f_c\prima(\id_{\bar v}):=\emptyset$, $f_c\prima(X_{\bar v}):=f_c(X_{\bar v})$, $f_c\prima(Y_{\bar v}):=f_c(Y_{\bar v})$ and $f_c\prima(Z_{\bar v})$ is defined in analogy with $f_c(X_{\bar v})$. The new mappings $f_c\prima$ are not group morphisms, but they do keep the good properties of the $f_c$ mappings listed above. And they satisfy $|O|\geq |f_c\prima(O)|$, which immediately leads to the bound $d\geq d_L$ with $d_L$ the minimal length, in terms of the number of edges, among nontrivial closed loops in $\Lambda$.

\subsection{Syndrome extraction}

As indicated in section \ref{sec:topological_memory}, in a topological quantum memory one has to keep track of errors by performing round after round of syndrome extraction. This raises the question of how fast and simply the stabilizer generators of a code $\code_\Lambda$ can be measured. The faster the measurements the less errors will be produced in the meanwhile, and the simpler they are the less faulty gates they will involve. Of course, what fast and simple really mean will depend on particular implementations, that is, in the basic operations at our disposal.

\begin{figure}
\psfrag{1}{1}
\psfrag{2}{2}
\psfrag{3}{3}
\psfrag{4}{4}
\psfrag{5}{5}
\psfrag{6}{6}
 \includegraphics[width=5cm]{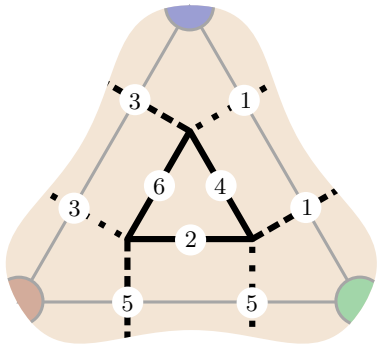}
 \caption{
The proposed ordering for the measurements of the edge operators. It does not depend on the particular geometry of the lattice $\Lambda$ because it is dictated by the coloring of its vertices.}
 \label{fig:measurements}
\end{figure}

To keep the discussion general, take gauge generator measurements to be the basic components of the syndrome extraction. At each time step the measurement of any subset of generators $\sset{G_i}$ is allowed as long as each physical qubit only appears in one of the $G_i$. Then, in any code $\code_\Lambda$  it is possible to cyclically measure all the stabilizer generators by performing six rounds of measurements. The time step at which each generator is to be measured is indicated in Fig.~\ref{fig:measurements}. Notice that $Z$-edges are measured at even times and $X$- and $Y$-edges at odd times. From the time steps 1-3 the eigenvalue for operators $S_v^c$ at blue vertices are obtained, from the steps 3-5 those for red vertices, and from the steps 5, 6 and 1 (this last one in the subsequent cycle) those for green vertices. It is not clear whether this number of time steps is optimal, since in principle 4 or 5 could be enough. As a comparison, the number of steps needed for Bacon-Shor codes is 4. In this sense the 6 steps are not bad, taking into account that the codes $\code_\Lambda$ do not benefit from the the separation of gauge and stabilizer generators into $X$-type and $Z$-type, as Bacon-Shor codes do.

\subsection{The problem of boundaries}\label{sec:subsystem_boundaries}

This section shows why it is not possible to introduce boundaries with the properties discussed in section \ref{sec:boundaries}. This has important practical consequences, since there is no other known way to introduce a nontrivial topology in a completely planar code. Notice however that we can always flatten a manifold to get a `planar' code, at the price of doubling the density of physical qubits in the surface. Also, the absence of boundaries makes less practical the use of code deformations, although they are still possible, as shown in section \ref{sec:subsystem_deformation}. In any case, this leaves open the question of whether other kind of interesting boundaries can be introduced.

\begin{figure}
\psfrag{a}{$\gamma$}
\psfrag{b}{$\gamma\prima$}
 \includegraphics[width=6cm]{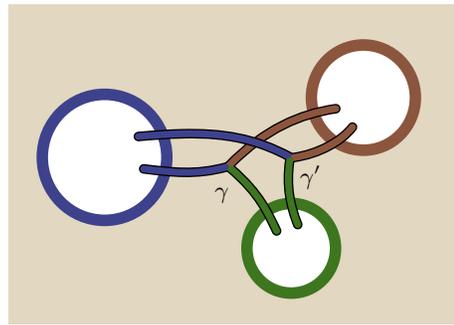}
 \caption{A hypothetical geometry for a $\code_\Lambda$ code with three holes of different colors. According to the properties of boundaries, the string operators $a$ and $b$ are equivalent up to stabilizer elements. This is a contradiction because due to the way they cross they anticommute.}
 \label{fig:no_boundaries}
\end{figure}

The existence of boundaries leads to the following contradiction. According to the properties listed in section \ref{sec:boundaries}, there are three potential kinds of boundaries, one per color. Each of them only allows strings of its color to end on it. Clearly either all the boundaries can be constructed or none of them. Thus, suppose that the three of them are allowed and consider a geometry as the one in Fig.~\ref{fig:no_boundaries}, with three holes, one of each color. Take a string-net $\gamma$ that connects the three holes, as in the figure, and deform it to another string-net $\gamma\prima$.  It follows from the properties of string operators and boundaries that $O_\gamma O_{\gamma\prima}\in \stab\prima$, but also that $\sset{O_\gamma,O_{\gamma\prima}}=0$ since they cross at a single point where they have different color. This is not possible.

In section \ref{sec:subsystem_string_operators} it was noted that the string label group $L\prima$ is the same in surface codes and the subsystem codes. Since according to section \ref{sec:boundaries} the set of allowed boundaries is dictated by $L\prima$, it could be expected that surface codes would neither have boundaries. However, this is not the case: two kind of boundaries can be constructed in surface codes \cite{Dennis_TQM}. The point is that there is a key difference between the two families of codes: in surface codes the three types of strings are not equivalent in any sense, so that the previous reasoning is not valid.

At a deeper level, this difference between the codes has its origin in the difference between the corresponding topological orders. Indeed, $L\prima$ does not encode all the information about the properties of anyons. In surface codes two of the quasiparticle types are bosons, and the third a fermion \cite{Kitaev_nonAbelian}, whereas in the subsystem codes the three are fermions \cite{Bombin_2body}. The connection between anyon condensation and boundaries is thus crucial: nice boundaries cannot be introduced in these topological subsystem codes because all the string operators are related to fermions, which cannot condense.

\subsection{Code deformation}\label{sec:subsystem_deformation}

This section explores the potential of code deformations in the topological subsystem codes $\code_\Lambda$. We show how initializations and measurements of individual logical qubits in the $X$ and $Z$ basis are possible through certain topology-changing processes on the manifold. And also, that CNot and Hadamard gates can be in principle implemented through continuous deformations of the manifold, but not in a practical way.

\begin{figure}
\psfrag{(a)}{(a)}
\psfrag{(b)}{(b)}
 \includegraphics[width=8cm]{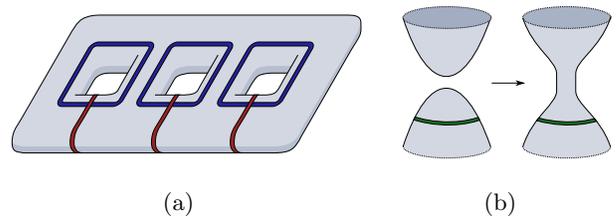}
 \caption{
(a) A sphere with $h$ holes can encode $2h$ qubits, but we choose to encode just $h$. The logical $\hat X_i, \hat Z_i$ operators correspond to the strings in the figure. Red strings give $X$'s and blue strings $Z$'s. (b) When the topology of the surface changes as indicated here two qubits are introduced in the code. They are initialized in a fixed way. In particular, the string operator in the figure is a boundary before the deformation takes place and thus has a fixed value. This value is not changed by the deformation because it occurs in a different part of the code.}
 \label{fig:deformation_A}
\end{figure}

To begin with, a manifold and a set of logical operators $\hat X_1, \hat Z_1, \cdots \hat X_k, \hat Z_k$ must be selected. We choose a $h$-torus, that is, a sphere with $h$ holes. Codes $\code_\Lambda$ on such a manifold provide $2h$ logical qubits, but only $h$ of them will be used, with the choice of logical operators indicated in Fig.~\ref{fig:deformation_A}(a). The rest of logical qubits are considered gauge qubits.

When a new handle is introduced, a logical qubit is created and initialized in a definite way, see Fig.~\ref{fig:deformation_A}(b). In the figure the two surfaces are supposed to be already connected, so that a handle is really created. There are two ways to introduce a new handle in a surface such as that of Fig.~\ref{fig:deformation_A}(a), depending on whether the process of Fig.~\ref{fig:deformation_A}(b) occurs inside' or `outside' the surface. In the former case the new qubit is initialized in a $\hat Z$ eigenstate and in the latter in a $\hat X$ eigenstate.  Whether the initialization occurs in the $Z$ or $X$ basis depends on which of the two string operators of the new qubit was a boundary initially. This operator has its eigenvalue fixed before the deformation occurs, and during the process it is topologically protected at all times\cite{Bombin_deformation}. The particular sign of the eigenstate depends on the arbitrary sign choices for the logical operators and $\stab_\Lambda$. If the initialization process is reverted it yields a measurement in the corresponding basis\cite{Bombin_deformation}.

It is always possible to detach a qubit, a torus, from the rest of the code. This does not involves any measurement, because the strings running along the cutting line are boundaries \cite{Bombin_deformation}. Similarly, there is no problem in attaching a torus to the code to add a logical qubit. But once a logical qubit is isolated, it can undergo code deformations independently. Consider a mapping that exchanges the two principal cycles of the torus and shifts the lattice a bit if necessary to adjust the color correspondence, for example by rotating the torus. Such a mapping can exchange $\hat X$ and $\hat Z$ operators, which amounts to a Hadamard gate. There exist an important drawback, though. This deformation cannot be realized in 3D without producing self-intersections of the surface. Still, it is conceptually interesting that the Hadamard gate can be obtained from purely geometric code deformations because this is not possible in surface or color codes, where $X$ and $Z$ type operators correspond to different type of string operators and a transversal Hadamard gate must be added to the picture\cite{Dennis_TQM}. Because color is just a matter of location in the lattice, strings of different colors are equivalent up to lattice translations. This is in essence what makes it possible the geometric implementation of the Hadamard gate.

\begin{figure}
\psfrag{(a)}{(a)}
\psfrag{(b)}{(b)}
\psfrag{(c)}{(c)}
 \includegraphics[width=8cm]{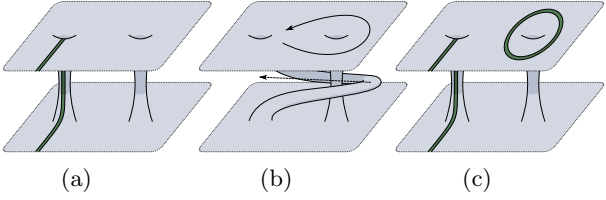}
 \caption{
The deformation that produces a controlled phase gate. (a)The code before the deformation takes place and a particular string. (b) The deformation moves one of the `holes' in the top part around the other, as indicate by the solid line with an arrow. To recover the original shape, as indicated by the dashed line, the two `tubes'  have to overlap unavoidably. (c) After the deformation, the string operator has been mapped to the product of these two string operators.}
 \label{fig:deformation_B}
\end{figure}

A controlled phase gate $\id - \half (1-Z_i)(1-Z_j)$ on a pair of logical qubits $i,j$ can be implemented through a `continuous' deformation of the code. The process is indicated in Fig.~\ref{fig:deformation_B}. It follows from the way in which the logical operatorators evolve \cite{Bombin_deformation} that the complete process amount to a controlled phase gate, up to some signs in the final logical operators that depend on the choice of $\stab_\Lambda$. A CNot gate can then be obtained by composing this gate with Hadamard gates. But again, such a code deformation requires to overlap the surface of the code with itself, see Fig.~\ref{fig:deformation_B}(b).

\section{Statistical physics of error correction}

In \cite{Dennis_TQM} an interesting connection between error correction thresholds for surface codes and phase transitions in 2D random bond Ising models was developed. Similar mappings exist also for color codes \cite{Katzgraber_3body}, in this case to 2D random 3-body Ising models. In both cases, the CSS structure of these codes is an important ingredient in the constructions: they are subspace codes with $\stab = \stab_X\stab_Z$ in such a way that $\stab_\sigma$ is generated by products of $\sigma$ operators, $\sigma = X, Z$. To take full profit of this, the noise channel for each qubit must be a composition of a bit-flip channel $\channel_{\text {bf}}(p):=\sset{(1-p)^\half \id, p^\half X}$ and a phase flip channel $\channel_{\text{pf} }(p):=\sset{(1-p)^\half \id, p^\half Z}$.

There are two main obstacles to construct a similar mapping for the codes $C_\Lambda$. The first is that they are subsystem codes rather than subspace codes. The second, that the gauge group cannot be separated in a $X$ and a $Z$ part. As we show below, both can be overcome.

\subsection{Mapping to a statistical model}

Rather than directly considering the codes $\code_\Lambda$, this section deals with the general mapping from any given stabilizer subsystem code to a suitable classical statistical model. For simplicity, each qubit in the code is supposed to be subject to a depolarizing channel $\channel_{\text{dep}}(p):=\sset{(1-p)^\half \id, (p/3)^\half X, (p/3)^\half Y, (p/3)^\half Z}$, with $p$ the error probability, but more general channels are possible within the same framework.

To build the classical Hamiltonian model, the first step is the choice of a set of generators $\sset{G_i}_{i=1}^l$ of $\gauge/\langle i \id\rangle$. These generators can be captured in a collection of numbers $g^\sigma_{ij}=0,1$ defined by
\begin{equation}\label{g_ij}
G_i\sigma_j=(-1)^{g_{ij}^\sigma}\sigma_jG_i,
\end{equation}
with $\sigma = X,Y,Z$, $i=1,\dots, l$, $j=1,\dots, n$. Attach a classical Ising spin $s_i=\pm 1$ to each of the generators $G_i$. The family of Hamiltonians of interest is
\begin{equation}\label{Hamiltonian}
H_{\tau} (s) := -    J \sum_{\sigma=X,Y,Z} \,\sum_{j=1}^{n}\, \,\tau_j^\sigma \, \prod_{i=1}^l \,s_i^{g_{ij}^\sigma},
\end{equation}
with parameters $\tau_j^\sigma= \pm 1$ such that $\tau_j^X\tau_j^Y\tau_j^Z=1$. 
The coupling $J>0$ is introduced to follow conventions. Notice that codes with local gauge generators give rise to local Hamiltonian models. Since $g_{ij}^X+g_{ij}^Y+ g_{ij}^Z=0 \mod 2$ the Hamiltonian \eqref{Hamiltonian} can be rewritten as 
\begin{equation}\label{Hamiltonian_alt}
H_{\tau} (s) = n-  \sum_{j} (1+ \tau_j^X \prod_i s_i^{g_{ij}^X})(1+ \tau_j^Y \prod_i s_i^{g_{ij}^Y}).
\end{equation}
The partition function for these Hamiltonians is
\begin{equation}\label{partition}
Z(K,\tau) = \sum_{s} e^{-\beta H_{\tau}(s)}
\end{equation}
with $K:=\beta J$ and $\beta$ the inverse temperature.

The goal is to express the class probabilities $p(\bar E)$, $E\in\pauli_n$, in terms of the partition function \eqref{partition} for a suitable $\tau$. Let $\tau=\tau_E$ be such that 
\begin{equation}\label{tau_E}
E\propto \bigotimes_j X^{\frac {1-\tau_j^Y}2} Y^{\frac{1-\tau_j^X}2}.
\end{equation}
Similarly,  for each $G\in \gauge$ choose any $s=s_G$ such that
 \begin{equation}\label{mu_G}
G\propto G_1^{\frac {1-s_1} 2}\cdots G_l^{\frac {1-s_l} 2}.
\end{equation}
We write $s\primas=s\prima s$ if $s\primas_j =s\prima_j s_j$. Then $s_Gs_{G\prima}=s_{GG\prima}$ and for any spin configuration $s$, $E\in\pauli_n$ and $G\in \gauge$ it can be checked that
\begin{equation}\label{A_property}
H_{\tau_{EG}}(s)=H_{\tau_E}(s_G s),
\end{equation}
In the depolarizing channel the probability for a Pauli error $E$ is $p(E)=(p/3)^{|E|}(1-p)^{n-|E|}$. It may be written as
\begin{equation}\label{A_probability}
p(E)= c_p^{-n} e^{-\beta_p H_{\tau_E}(s_\id)},
\end{equation}
where $c_p:= e^{3K_p} +3e^{-K_p}$ and $\beta_p:=K_p/ J$ with 
\begin{equation}\label{Nishimori}
3e^{-4K_p} := \frac p {1-p}.
\end{equation}
The desired connection follows from \eqref{A_property} and \eqref{A_probability}, which give
\begin{equation}\label{probability_partition}
p(\bar E)= \sum_{G\in \gauge} p(EG)= \frac 1 {2^{w} c_p^{n}}Z(K_p,\tau_E),
\end{equation}
where $w$ is the number of redundant generators of $\gauge$, that is, $w=l-l\prima$ with $l\prima$ the rank of $\gauge/\langle i\id\rangle$.

\subsection{CSS-like codes}\label{sec:CSS_stat}

To connect the results of the previous section with the works \cite{Dennis_TQM, Wang_topoStat, Katzgraber_3body} CSS codes must be considered. These are codes with $\gauge = \langle i\id\rangle \gauge_X\gauge_Z$ for some $\gauge_\sigma$ generated by products of $\sigma$ operators, $\sigma = X, Z$. And, instead of a depolarizing channel, the noisy channel must take the form $\channel = \channel_{\text{bf} }(p)\circ\channel_{\text{pf} }(p\prima)$. This allows to treat $X$ and $Z$ errors independently \cite{Dennis_TQM}. Here we consider the case of bit flip errors, phase flip errors are analogous.

The construction is similar to the one in the previous section. It starts with the choice of generators $\gauge_{X}=\langle G_i\rangle_{i=1}^l$. The relevant Hamiltonians read
\begin{equation}\label{Hamiltonian_p}
H_\tau\prima := - J \sum_{j=1}^{n}\, \tau_j \, \prod_{i=1}^l \,s_i^{g_{ij}},
\end{equation}
where $\tau_j:=\tau_j^Z$ and $g_{ij}:=g_{ij}^Z$. The probability of an error $E$ that is a product of $X$ operators is
\begin{equation}\label{probability_partition_p}
p(\tilde E):= \sum_{G\in \gauge_X} p(EG)= \frac 1 {2^{w} (2\cosh K_p\prima)^{n}}Z(K_p\prima,\tau_E),
\end{equation}
where $w$ is the number of redundant generators of $\gauge_X$ and
\begin{equation}\label{Nishimori_p}
e^{-2K_p\prima} := \frac p {1-p}
\end{equation}
defines the Nishimori temperature \cite{Nishimori_statInfo}.

Bacon-Shor codes provides an example of gauge CSS-like codes. With the above procedure, they yield models that amounts to several copies of the 1D Ising model.

\subsection{Symmetries}

Interestingly, the redundancy of the generators of $\gauge$ is directly connected to the symmetries of the Hamiltonians \eqref{Hamiltonian}. Suppose that the generators are subject to a constraint of the form 
\begin{equation}\label{constraint_generators}
\prod_{i\in I} G_i \propto \id
\end{equation}
for some set of indices $I$. 
Then \eqref{A_property} gives
\begin{equation}\label{symmetry}
H_\tau (s)= H_\tau (s\prod_{i\in I} s_{G_i}).
\end{equation}
In other words, making the most natural choice for the $s_{G_i}$ it follows that the Hamiltonian is invariant under the transformation
\begin{equation}\label{symmetry_transformation}
s_i\longrightarrow s_i\prima= \begin{cases} -s_i, &i\in I, \\ s_i, &i\nin I. \end{cases}
\end{equation}
Thus, global constraints lead to global symmetries and local constraints to local symmetries. 

As a particular example, consider surface codes, which are mapped to Ising models \cite{Dennis_TQM}. In these codes the product of all $X$-type stabilizers equals the identity, producing a symmetry that is simply the global $\Z_2$ symmetry of the Ising model.

\subsection{Error correction and free energy}

We now put equation \eqref{probability_partition} to use in the error correction framework of section \ref{sec:error_correction}. Recall that, after the syndrome has been measured, one has to find the most probable class of errors among several candidates $\bar E_i := \bar E\bar D_i$. This amounts to compare the probabilities $p(\bar E_i)$ or, alternatively, the quantities $Z(K_p,\tau_{E_i})$. And to do this, it is enough to know the free energy differences \cite{Dennis_TQM}
\begin{equation}\label{free_energy}
\Delta_i(K_p, \tau_E):=  \beta F (K_p, \tau_{E_i}) - \beta F(K_p,\tau_E),
\end{equation}
where $F(K, \tau) = - T\log Z(K, \tau)$ is the free energy of a given interaction configuration $\tau$. For example, in the Ising models that appear for 2D surface and color codes these are domain wall free energies. 

In practice, the computation of \eqref{free_energy} may be difficult. In this regard, it has been suggested \cite{Dennis_TQM}, in the context of surface codes, that in the absence of glassy behavior the computation of \eqref{free_energy} should be manageable, and in \cite{Wang_topoStat} a possible approach was sketched.

\subsection{Error threshold and phase transition}

In surfaces codes there exists an error probability $p_c$, the error threshold, such that the asymptotic value of the success probability $p_0$,  in the limit of large code instances, is one for $p<p_c$ and  $1/4^k$ for $p>p_c$ \cite{Wang_topoStat}. This is directly connected to an order-disorder phase transition in a model with random interactions. An analogous transition is observed for the random model that corresponds to color codes \cite{Katzgraber_3body, Ohzeki_CC, Landahl_CC}. It is then natural to expect a similar connection in other topological codes, as we describe next.

Consider a random statistical model with Hamiltonian \eqref{Hamiltonian} in which the parameter $\tau$ is a quenched random variable. That is, $\tau$ is random but not subject to thermal fluctuations. The probability distribution $p(\tau)$ is such that the signs of $\tau_i^{\sigma}$ and $\tau_j^{\sigma\prima}$ are independent if $i\neq j$. For each $i$, the case $\tau_i^X = \tau_i^Y= 1$ has probability $1-p$ and the other cases have probability $p/3$ each. In other words, if $\tau = \tau_E$ then $p(\tau)=p(E)$ with $p(E)$ given by the depolarizing channel $\channel_{\text{dep}}(p)$. 

In thermal equilibrium the model has two parameters, the temperature $T$ and the probability $p$. For the mapping only a particular line in the $p$-$T$ plane is relevant, the Nishimori line \cite{Nishimori_statInfo}, given by the condition $K=K_p$ that has its origin in \eqref{A_probability}. The error correction success probability in \eqref{success_prime} can be written in terms of this statistical model as follows:
\begin{equation}\label{success_stat}
p_0\prima=\left[\left ( 1+\sum_{i=2}^{4k} e^{-\Delta_i(K_p, \tau)} \right )^{-1}\right ]_{K_p},
\end{equation}
where $[ \cdot  ]_{K_p} := \sum_{\tau} p(\tau) \,\cdot$ denotes the average over the quenched variables. 

Suppose that the code has a threshold probability $p_c$ below which $p_0\prima\rightarrow 1$ in the limit of large codes. Then \cite{divergence}, in the random model the average of the free energy difference \eqref{free_energy} diverges with the system size, $[\Delta_i(K,\tau)]_{K_p}\rightarrow \infty$, for $p< p_c$ along the Nishimori line. This is exemplified \cite{Dennis_TQM} by surface codes and the corresponding random 2d Ising models, where $[\Delta_i(K,\tau)]_{K_p}$ is the domain wall free energy. It diverges with the system size below $p=p_c$ and attains some finite limit over the threshold, signaling an order-disorder phase transition at $p_c$. A similar behaviour can be expected for other topological codes. For 2D color codes this was shown in \cite{Katzgraber_3body}.

\subsection{The Hamiltonian model for $\code_\Lambda$ codes}

The above mapping can be immediately applied to the subsystem codes $\code_\Lambda$. Choose as generators of the gauge group the edge operators $O_e$, so that there is an Ising spin $s_e$ at each edge $e$. The Hamiltonian takes the form
\begin{equation}\label{Hamiltonian_lambda}
H_{\tau}^\Lambda (s) := - J \sum_{j=1}^n \tau_j^X s_{2}s_{3}s_{4} + \tau_j^Y s_1s_3s_4 +\tau_j^Z s_1s_2  ,
\end{equation}
where the sum runs over vertices and for each of them the Ising spins $s_1, s_2, s_3, s_4$ correspond respectively to the $X$, $Y$ and two $Z$ edges meeting at the vertex. 

The Hamiltonian \eqref{Hamiltonian_lambda} has a local symmetry at each triangle. In particular, flipping the three Ising spins of the triangle leaves $H_\tau^\Lambda$ invariant. This is so becuase the product of the three edge operators in the triangle equals the identity. There exist also a global $\Z_2\times\Z_2$ symmetry that follows from the global constraints in \eqref{constraints_stabilizer}.  The local constraints in \eqref{constraints_stabilizer} do not provide any symmetry as they are trivial in terms of the gauge generators.

\subsection{Faulty measurements}

The mapping considered up to know is only suitable if perfectly accurate quantum computations are allowed in error correction. This section generalizes it to include errors in the measurements of the stabilizer generators.

Following \cite{Dennis_TQM}, take as a goal the `indefinite' preservation of the content of a quantum memory. Time is divided in discrete steps. At each time step, the memory suffers errors and at the same time the stabilizer generators are imperfectly measured. Then if from the history of measurements one can correctly infer the actual history of errors, up to a suitable equivalence, the memory is safe. 

The results in \cite{Dennis_TQM, Wang_topoStat} show that for surface codes there exists of a noise threshold below which long time storage is possible for sufficiently large codes. The same behavior can be expected for other topological codes, but the construction of a suitable random statistical model for each code is required first. Here we generalize the construction of \cite{Dennis_TQM} to subsystem codes and depolarizing channels.

\subsubsection{Depolarizing channel}

Consider first the case of a depolarizing channel $\channel_{\text{dep}}(p)$ occurring for each physical qubit between each round of measurements. We adopt the convention that at a given time $t$ first errors occur and then faulty measurements are performed.

Recall that in the mapping of error correction to a statistical model errors were mapped to interactions through the $\tau_j^\sigma$, see \eqref{tau_E}. The new elements here are time and faulty measurements. Since errors can occur at different time steps $t$, a time label must be added to the $\tau_j^\sigma$'s to get the collection of signs $\tau=(\tau_{jt}^\sigma)$, subject as before to the constraints $\tau_{jt}^X\tau_{jt}^Y\tau_{jt}^Z=1$. To represent errors in the measurements of stabilizers, first a set $\sset{S_k}_{k=1}^m$ of generators of $\stab$ to be measured at each time step $t$ must be chosen. Attach to them a collection of signs $\kappa_{kt} = \pm 1$. The correct (wrong) measurement  of the $i$-generator at time $t$ corresponds to $\kappa_{kt}=1$ ($\kappa_{kt}=-1$). In the statistical model the $\tau$ and $\kappa$ are quenched variables. $\tau$ follows the same distribution as before, dependent on the probability $p$, and each $\kappa_{kt}$ is independent an takes value $-1$ with probability $q$. For this to make sense under the mapping, errors in the measurements must occur independently with a fixed probability $q$. This will not be true in most settings. Still, it is a useful assumption because knowing the correlations between errors can only improve error correction. In analogy with the $g_{ij}^\sigma$ defined above, the stabilizer generators are captured in a collection of numbers $h^\sigma_{kj}=0,1$ defined by
\begin{equation}\label{g_ij}
S_k\sigma_j=(-1)^{h_{kj}^\sigma}\sigma_jS_k,
\end{equation}
with $\sigma = X,Y,Z$, $k=1,\dots, m$, $j=1,\dots, n$.

Recall also that in the original mapping gauge generators $G_i$ were mapped to Ising spins $s_i$. The reason for this was that gauge generators play the role of basic equivalences between errors. Now for errors that occur at the same time $t$ these kind of equivalence happens again, represented by spins $s_{it}$. But in addition there is an equivalence between errors that involves errors at different time and measurement errors. If at times $t$ an $t+1$ a given error occurs and the measurements at time $t$ of the stabilizers that would detect the error fail, then these errors altogether go unnoticed but produce no harm. Thus, two collections of errors that differ only by such an event should be considered equivalent. Therefore, Ising spins that represent this equivalence are necessary. This can be achieved by attaching two Ising spins $s^X_{jt},  s^Y_{jt}$ to the $t$-th time step and $j$-th qubit.  The Hamiltonians are
\begin{align}\label{Hamiltonian_faulty}
H_{\tau, \kappa} (s) := &- J  \sum_\sigma \sum_j \sum_t \,\tau_{jt}^\sigma\, s_{j(t-1)}^\sigma \, s_{jt}^\sigma\,\prod_i \,s_{it}^{g_{ij}^\sigma} \,\,+\nonumber\\ 
&- K  \sum_k \sum_t \,\kappa_{kt}\, \prod_j \prod_\sigma \,(s_{jt}^\sigma)^{h^\sigma_{kj}},
\end{align}
where $s^Z_{jt}:=s^X_{jt} s^Y_{jt}$ and the range of values of the different indices should be clear from the context. In order to recover the probability of a given set of errors from the partition function the relations
\begin{equation}\label{Nishimori_faulty}
3e^{-4\beta J} = \frac p {1-p},\qquad e^{-2\beta K} = \frac q {1-q}
\end{equation}
must hold.

For each time step $t$, the Hamiltonians \eqref{Hamiltonian_faulty} keep the symmetries \eqref{symmetry_transformation}. In addition, there is a symmetry for each gauge generator $G_{i\prima}$ and time $t\prima$. Namely,
\begin{align}\label{symmetry_transformation_faulty}
s_{jt}^\sigma&\longrightarrow s_{jt}^{\prime\sigma}=\begin{cases} (-1)^{g_{i\prima j}^\sigma} s_{jt}^\sigma, &t=t\prima,\\ s_{jt}^\sigma, &t\neq t\prima,\end{cases}\nonumber\\
s_{it}&\longrightarrow s_{it}\prima= \begin{cases} -s_{it}, &i=i\prima, \,t=t\prima,t\prima+1, \\ s_{it}, &\text{otherwise.} \end{cases}
\end{align}
Therefore, local gauge generators give rise to a (random) gauge model.

\subsubsection{Bit flip channel}

Finally, consider the simpler case of a bit flip channel $\channel_{\text{bf}}(p)$ in a CSS-like code. As noted above, the case of a phase flip channel is analogous and if both channels happen consecutively they can be treated independently. 

The construction is an extension of the one in section \ref{sec:CSS_stat}. The $\tau_{j}$'s and $s_i$ are respectively replaced by the signs $\tau_{jt}$ and the Ising spins $s_{it}$. Given a choice of generators $\stab_{Z}=\langle S_k\rangle$ to be measured at each time step, there is a corresponding collection of signs $\kappa_{kj}$. The generators take the form $S_k=\pm \bigotimes_j Z_j^{h_{kj}}$ for some $h_{kj}=0,1$. There is also an Ising spin $\hat s_{jt}$ for each physical qubit $j$ and time step $t$. The Hamiltonians read
\begin{align}\label{Hamiltonian_faulty_p}
H_{\tau, \kappa}\prima (s) := &- J  \sum_j \sum_t \,\tau_{jt}\, \hat s_{j(t-1)} \, \hat s_{jt}\,\prod_i \,s_{it}^{g_{ij}} \,\,+\nonumber\\ 
&- K  \sum_k \sum_t \,\kappa_{kt}\, \prod_j  \,\hat s_{jt}^{h_{kj}}.
\end{align}
Instead of \eqref{Nishimori_faulty}, the right conditions are now
\begin{equation}\label{Nishimori_faulty_p}
e^{-2\beta J} = \frac p {1-p},\qquad e^{-2\beta K} = \frac q {1-q}.
\end{equation}
The analog of \eqref{symmetry_transformation_faulty} is
\begin{align}\label{symmetry_transformation_faulty_p}
\hat s_{jt}&\longrightarrow \hat s_{jt}\prima=\begin{cases} (-1)^{g_{i\prima j}} s_{jt}, &t=t\prima,\\ \hat s_{jt}, &t\neq t\prima,\end{cases}\nonumber\\
s_{it}&\longrightarrow s_{it}\prima= \begin{cases} -s_{it}, &i=i\prima, \,t=t\prima,t\prima+1, \\ s_{it}, &\text{otherwise.} \end{cases}
\end{align}

\section{Conclusions}

Topological codes are intrinsically local, and gauge or subsystem codes can have interesting locality properties. In this paper we have introduced a family of topological subsystem codes, thus putting together the two concepts. The gauge group of the code is generated by 2-local operators, which compares well with surface or color codes that have at least 4-local and 6-local generators, respectively. In particular, the measurement of these 2-local operators is enough to recover the error syndrome.

We have argued that these codes do not allow the introduction of boundaries with nice properties, which motivates further research. There are probably interesting topological codes still to be discovered. One could look for example for subsystem codes with nice boundaries or with interesting transversality properties as those found in color codes.

We have also explored a general connection between error correction in subsystem codes and statistical physics. The connection is specially meaningful in the case of topological codes, where the error threshold maps to a phase transition in the corresponding statistical model. There is a lot of work to do in this direction. For example, the computation, probably numerically, of the error threshold of the topological subsystem codes presented here.

\section*{Acknowledgements}

I would like to acknowledge useful and encouraging discussions with D. Gottesman, J. Preskill, A. Kitaev, A. Landhal and H. Katzgraber. I also thank M.A. Martin-Delgado for his comments on a preliminary version. Part of this work was carried out during invited stays at California Institute of Technology, University of New Mexico and Texas A\&M University. I acknowledge financial support from a PFI grant of EJ-GV, DGS grants under contract FIS2006-04885, and the ESF INSTANS 2005-10.

\appendix



\section{Structure of $N(\gauge)$}\label{app:gauge}

This appendix is a complement to section \ref{sec:subsystem_string_operators}, and uses the same notation.

A color code $\code_\Lambda^\rc$ \cite{Bombin_CC2d} can be obtained from a lattice $\Lambda$ with the properties enumerated in section \ref{sec:subsystem_lattice}. The construction is the following. First, there is one qubit per triangle, so that the relevant Pauli group is $\pauli_{|F|}$. Given a collection of triangles $T=\sset{\tau_i}$, set $X_T:=\bigotimes_i X_{\tau_i}$, $Z_T:=\bigotimes_i Z_{\tau_i}$. If each vertex $v\in V$ is identified with the set of triangles meeting at $v$, the stabilizer for the color code is $\stab_\Lambda^\rc := \stab_X \stab_Z$ with $\stab_X := \langle X_v \rangle_v$, $\stab_Z := \langle Z_v \rangle_v$. Let $\sset {T_i}$ be the collection of those sets of triangles that have an even number of triangles meeting at each vertex. Then $N(\stab_\Lambda^\rc)= \langle i\id\rangle N_XN_Z$ with $N_X:=\langle X_{T_i}\rangle_i$, $N_Z:=\langle Z_{T_i}\rangle_i$. 

Next, consider the morphism $\funcion f {N(\gauge_\Lambda)} {N_X}$ such that $f(O)=X_{T_\gamma}$ for any subgraph $\gamma$ of $\bar\Lambda$ and $O\in N(\gauge_\Lambda)$ such that $O\propto O_\gamma$. The kernel of $f$ is formed by those operators that only involve $Z$-s, not $X$-s or $Y$-s. That is, $\ker f = \langle i\id \rangle  \langle O_v^{c_v}\rangle_v\subset \mathcal S\prima$, where $c_v$ is the color of $v$ in $\Lambda$. Since $f(O_v^c)=X_v$ for $c\neq c_v$ , $f[\stab\prima_\Lambda]=\stab_X$ and thus $\stab\prima_\Lambda/\ker f \simeq  \stab_X$. This implies that there are no other constraints for the generators of $\stab\prima_\Lambda$ apart from the ones in \eqref{constraints_stabilizer}, because exactly two of the generators $\sset{X_v}$ of $\stab_X$ are unnecessary \cite{Bombin_CC2d}. Finally, it is easy to check that for any string operator $X_T\in N_X$, as described in \cite{Bombin_CC2d}, there exists a string-like graph $\gamma$ such that $f(O_\gamma)=X_T$, so that $f$ is onto and $N(\gauge_\Lambda)/\stab\prima_\Lambda\simeq N_X/\stab_X$. Then the properties of the string operators in $\code_\Lambda$ are consequences of those for string operators in $\code_\Lambda^\rc$. This is in particular true regarding the generating set and the composition rules, but not for commutation rules, which have to be worked out separately.



\begin{thebibliography}{99}
\bibitem{divergence}
Let us show that if $p_0\prima$ in \eqref{success_stat} is $1$ in a certain limit, then $[\Delta_i(K_p,\tau)]_{K_p}\rightarrow \infty$ for $i\neq 1$. Given $r> 0$, let $T$ be the set of those $\tau$'s with $\Delta_i(K_p,\tau)< r$ for some $i\neq 1$, and set $p_r:= \sum_{T} p(\tau)$. Then $p_0\prima \leq (1-p_r)+p_r (1+e^{-r})^{-1}=1-p_r (1+e^r)^{-1}$. Thus for any $r$ we have $p_r\rightarrow 0$.
\end{thebibliography}
\end{document}